\documentclass[preprint,12pt]{elsarticle}

\usepackage{amssymb}

\usepackage{graphicx}
\usepackage[T2A]{fontenc}
\usepackage{amsmath,amssymb,epsfig}
\usepackage{comment}
\usepackage{color,subfigure}

\def\a {\alpha}
\def\ve {\varepsilon}

\def\b {\beta}
\def\g {\gamma} \def \G {\Gamma}
\def\d {\delta} \def\D {\Delta}
\def\l {\lambda} \def\L {\Lambda}
\def\m {\mu}
\def\n {\nu}
\def\p {\pi}

\def\s {\sigma}
\def\ph{\phi}

\def\ra{\rightarrow}
\def\dd{{\rm d}}

\def\({\left(}
\def\){\right)}
\def\[{\left[}
\def\]{\right]}

\def\bf{\textbf}

\def\beq {\begin{equation}}
\def\eq {\end{equation}}
\def\bea{\begin{eqnarray}}
\def\ea{\end{eqnarray}}

\def\({\left(}
\def\){\right)}
\def\[{\left[}
\def\]{\right]}

\newcommand{\notd}[1] { \setbox0=\hbox{$#1$}
\dimen0=\wd0   \setbox1=\hbox{/} \dimen1=\wd1  \ifdim\dimen0>\dimen1
 \rlap{\hbox to \dimen0{\hfil/\hfil}}  #1 \else \rlap{\hbox to \dimen1{\hfil$#1$\hfil}}  /  \fi  }

\journal{Physics Letters B}

\begin{document}

\begin{frontmatter}



\title{Analytic structure of $\phi^4$ theory using light-by-light sum rules}


\author[lab1,lab2,lab3]{V.~Pauk}
\author[lab1,lab2]{V.~Pascalutsa}
\author[lab1,lab2]{M.~Vanderhaeghen}

\address[lab1]{PRISMA Cluster of Excellence, Johannes Gutenberg-Universit\"at,  Mainz, Germany}
\address[lab2]{Institut f\"ur Kernphysik, Johannes Gutenberg-Universit\"at,  Mainz, Germany}
\address[lab3]{Department of Physics, Taras Shevchenko National University of Kyiv, Ukraine}

\begin{abstract}
We apply a sum rule for the forward light-by-light scattering process 
within the context of the $\phi^4$ quantum field theory. 
As a consequence of the sum rule a stringent causality criterion is presented and the resulting constraints are studied within a particular resummation of graphs. 
Such resummation is demonstrated to be consistent with the sum rule to all orders of perturbation theory. We furthermore show the appearance of particular non-perturbative solutions within such approximation to be a necessary requirement of the sum rule. For a range of values of the coupling constant, these solutions manifest themselves as a physical 
bound state and a $K$-matrix pole. For another domain however, they appear as tachyon solutions, showing 
the inconsistency of the approximation in this region. 
\end{abstract}


\end{frontmatter}













\section{Introduction}

Sum rules provide a powerful tool to study relativistic quantum field theories, and apply 
also outside the regime where perturbative expansions hold.  
As sum rules are consequences of such general principles as analyticity and unitarity, they allow to establish rigorous relations between  physical observables, even when the underlying theory is non-perturbative in nature and cannot be solved exactly. 

In recent works~\cite{Pascalutsa:2010sj,Pascalutsa:2012pr}, we have derived three sum rules for the 
low-energy forward light-by-light scattering process. These $\gamma \gamma$-sum rules are non-perturbative in origin and demonstrate that the low and high-energy behaviors of the theory are related. We showed e.g. that a sum rule for the helicity-difference total cross-section of the photon-photon-fusion process, $\gamma \gamma \to X$, 
reveals in the hadron sector an intricate correlation between contributions of pseudoscalar and
tensor mesons. In the charm quark sector, the $\gamma \gamma$ sum rules reveal an interplay between $c \bar c$ bound states, 
open charmed meson continuum states, as well as exotic $c \bar c$ resonance states~\cite{Pascalutsa:2012pr}. 
Several experiments, primarily at $e^+ e^-$ collider facilities, are presently uncovering the rich spectroscopy of such systems, 
see e.g.~\cite{Nielsen:2009uh, Brambilla:2010cs} for some recent reviews.  

Besides its relevance to hadron phenomenology, we can use the $\gamma \gamma$-sum rules in the same  way in model field theories, where in the case of renormalizable models they can be applied perturbatively. When the conditions of applicability are fulfilled these sum rules were shown to hold in leading order calculations~\cite{Pascalutsa:2012pr}.  However, the realization of the causality constraints at higher orders as well as in the non-perturbative regime of quantum field theories is still an open issue. 

Studies of causality constraints on the basis of different sum rules were carried out in the past in a number of different contexts. Especially the realization of the well-known Gerasimov-Drell-Hearn sum rule~\cite{GDH} within perturbative field theory was analyzed 
for spin-1/2 targets at the lowest nontrivial order~\cite{wk} as well as at higher orders in QED~\cite{Dicus:2000cd}. In Refs.~\cite{Sucher:1974cw,Salam:1974bz} consequences of the sum rules within asymptotically free theories were considered. 
In more recent years, they have also been discussed within the context of quantum gravity~\cite{Goldberg:1999gc,Grigoryan:2012ew}.

In the present work, we are using light-by-light scattering sum rule as a tool to study causality constraints within a model field theory, the $\phi^4$ scalar theory. 
We consider a bubble-chain resummation and demonstrate it to be consistent with causality to all orders of perturbation theory. Furthermore, it is shown that the sum rule strictly defines the non-perturbative structure of the solutions which arise dynamically within this approximation. In a particular regime of the coupling constant the spectrum of solutions contains a dynamically generated bound state and a $K$-matrix pole. For another domain the solution possesses an unphysical pole with negative invariant mass being a direct sign of the inconsistency of the approximation.

The outline of this letter is as follows. 
In Sect.~\ref{sec2}, we compute the light-by-light scattering sum rule 
involving the helicity difference cross section for the $\gamma \gamma \to X$ process, within the $\phi^4$ scalar field theory at one-loop level. In Sect.~\ref{sec3}, we provide a calculation beyond the one-loop level in the ``bubble-chain" approximation. 
In Sect.~\ref{sec4} we discuss how causality imposes constraints on the solutions for different values of the renormalized self-interaction coupling constant of the $\phi^4$ theory. The summary and outlook are given in  Sect.~\ref{sec5}.

\section{One loop}
\label{sec2}

In this work, we will focuss on a sum rule for the forward light-by-light scattering, which involves the helicity-difference cross-section for real photons~\cite{Gerasimov:1973ja,Brodsky:1995fj,Pascalutsa:2010sj} and reads as:
\beq
\int\limits_{s_0}^\infty\dd s\,\frac{\D\s(s)}{s}=0,
\label{HDsr}
\eq
where $\D\s(s)=\s_2(s)-\s_0(s)$ is the total helicity-difference cross section of the two-photon fusion process $\gamma \gamma \to X$, 
with the  Mandelstam variable $s=(q_1+q_2)^2$, where $q_1$ and $q_2$ are the two photon 4-momenta, and $s_0$ is the lowest production threshold of the process. 

We will study the above sum rule in a particular model quantum field theory. 
We take one of the simplest examples: a self-interacting scalar field $\phi(x)$
with charge $e$ and mass $m$ as described
by the following Lagrangian density,
\beq
\mathcal{L}=(D^\m\ph)^* D_\m\ph-m^2\ph^*\ph + \frac{\l_0}4(\ph^*\ph)^2-\frac14F^{\m\n}F_{\m\n},
\label{lagrangian}
\eq
where  $\l_0$ is the self-interaction coupling constant, while
the covariant derivatives and electromagnetic field-strength tensor
are given as usual by $D_\m=\partial_\m+ie A_\m$ and $F_{\m\n} =
\partial_{\m} A_{\n} - \partial_{\n} A_{\m} $. 

We denote the helicity amplitudes
for the process $\g\g\to \phi\phi^*$ by ${M}_{++}$ and ${M}_{+-}$, where the subscripts indicate the photon helicities.  
 Given these amplitudes,
the cross section for total helicity-0 and 2 are found as:
\bea
\s_0(s) & = & \frac{\beta(s)}{32 \pi s}  \int_{-1}^1 d \cos \theta \, 
\left|{M}_{++}(s,\,\theta) \right|^2 , \\
\s_2(s) & = & \frac{\beta(s)}{32 \pi s}  \int_{-1}^1 d \cos \theta \, 
\left|{M}_{+-}(s,\,\theta) \right |^2,
\ea
where $\theta$ is the angle of one of the members of the $\phi\phi^*$ pair
w.r.t. the photon in the center-of-mass system, and 
where we introduced their relative velocity $\b$ as:
\beq
\b(s)=\sqrt{1-\frac{4m^2}s}.
\eq

\begin{figure}
\centering
  \includegraphics[width=12cm]{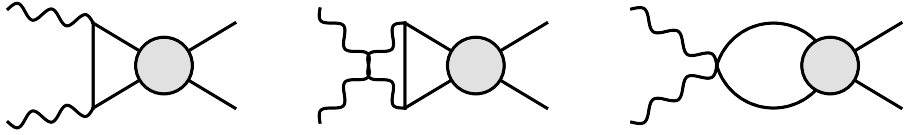}\\
  \caption{The contribution to the $\g\g$-fusion process within the $\phi^4$ field theory considered in this work. The solid lines denote the charged scalar fields.}
    \label{bubblechain}
\end{figure}

To leading order in $\tilde \l_0 \equiv \l_0 / (4 \pi)^2$ and in the fine-structure constant $\a \equiv e^2/4\pi$,
the helicity amplitudes are found to be:
\begin{subequations}
\bea
{M}_{++}(s, \theta) &=&  4 \pi \alpha \left\{ \frac{2 ( 1 - \beta^2) }{1 - \beta^2 \cos^2 \theta}
+ \tilde \lambda_0  \, 2  F(s) \right\}, 
\label{helampl} 
\\
{M}_{+-}(s, \theta) &=&  4 \pi \alpha  \frac{2 \beta^2 \sin^2 \theta }{1 - \beta^2 \cos^2 \theta}, 
\ea
\end{subequations}
where the first term in ${M}_{++}$ and the expression for  ${M}_{+-}$ correspond with the tree level amplitudes for  
$\gamma \gamma \to \phi \phi^*$ in scalar QED. The second term in ${M}_{++}$ describes the one-loop production 
process corresponding with Fig.~\ref{bubblechain}, with the grey blob denoting the four-particle 
amplitude evaluated to leading order in $\lambda_0$.  
Furthermore in Eq.~(\ref{helampl}), the (dimensionless) form factor $F(s)$ describing the transition of photons 
to a scalar pair at one-loop order is given explicitely by~:
\bea
F(s) = - 1 - \frac{m^2}s {\mathrm {Re}} \[ \ln \frac{1+\b(s)}{1-\b(s)} - i \p \]^2 
+ i \, \frac{2 \pi m^2}{s}  \theta(s-4m^2)   \ln \frac{1+\b(s)}{1-\b(s)} .
\ea
It is quite easy now to compute the cross sections for $\g\g\to \phi\phi^*$,
the result for the helicity difference cross-section is:
\beq
\D \s (s) = \D \s^{(\mathrm{tree})}(s) 
-  \a^2 \, \tilde \l_0 \,  \frac{8 \pi}{s}  \left( 1 - \beta^2(s) \right)  \, \mathrm{arctanh}\,\beta(s) \,\, \mathrm{Re}\, F(s),
\eq
where $\D \s^{(\mathrm{tree})}$ is the tree-level cross section 
in scalar QED (cf., e.g., Appendix in \cite{Pascalutsa:2012pr}).
The tree-level cross section weighted with $1/s$ integrates to zero by itself,
and it can easily be verified that
\beq
\int\limits_{4m^2}^\infty ds \, \frac{\mathrm{Re}\, F(s)}{ s^3}\, \mathrm{arctanh}\, \beta(s) = 0.
\eq
Hence, we have shown that the sum rule is obeyed at the one-loop level.

\section{Bubble-chain sum}
\label{sec3}

\begin{figure}[t]
\centering
\subfigure{\epsfig{figure=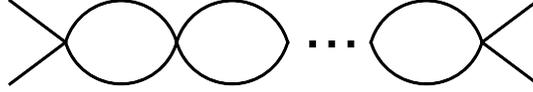,width=.42\textwidth}}
\caption{The bubble chain contribution entering the grey blob of Fig.~\ref{bubblechain}.}
\label{oneloop}
\end{figure}

A class of diagrams naturally arises when one analyzes how the sum rules are realized at higher orders in perturbation theory, and in fact 
beyond perturbation theory.  
In the following, we discuss the contribution of the bubble-chain type diagrams to the $\g\g$-fusion process as shown on Fig.~\ref{bubblechain}, where the shaded blob now denotes a bubble-chain contribution to the four-particle vertex, as shown in Fig.~\ref{oneloop}.
The bubble-chain approximation arises in many contexts,
but most notably as the leading large-$N$ result of the $\mathcal{O}(N)$ models.
The interest in such an approximation is due to the fact that it preserves much of the non-linear structure 
of the exact theory \cite{Coleman:1974jh,Abbott:1975bn}.

The bubble-chain only contributes to the helicity-0 amplitude ${M}_{++}$, with the 
$n$th-bubble contribution  being given by 
\beq
M^{(n+1)}_{++}(s)= 4 \pi \alpha \, \tilde \l_0 \, [\lambda_0 B(s)]^n \, 2 F(s) .
\label{nPTa}
\eq

In the dimensional regularization scheme the one-loop corrections to the four-particle vertex, corresponding with a single bubble in Fig.\ref{oneloop},  can be expressed through the scalar integral~:
\beq
B(s) \equiv - i \int \frac{d^d l}{(2 \pi)^d} \frac{1}{\left[ (p + l)^2 - m^2 + i \varepsilon \right] \left[ l^2 - m^2 + i \varepsilon  \right]},
\eq
with $p$ the total four-momentum, and $s = p^2$.  
This integral has the explicit form~:
\beq
B(s) = \frac{1}{(4\p)^2}\left\{ - L_\ve(\m^2) + 2 -\b(s)\,\ln\,\frac{1+\b(s)}{1-\b(s)} +i\p\b(s)\theta(s-4m^2) \right\}.
\eq
Here $L_\ve=-1/\ve+\g_E+\log(m^2/4\p\m^2)$ is the dimensional regularization factor and $\m$ is the corresponding dimensional regularization scale, $\g_E=-\G'(1)\simeq0.5772$ is Euler constant. 

\begin{figure}[h]
\centering
\includegraphics[scale=0.8]{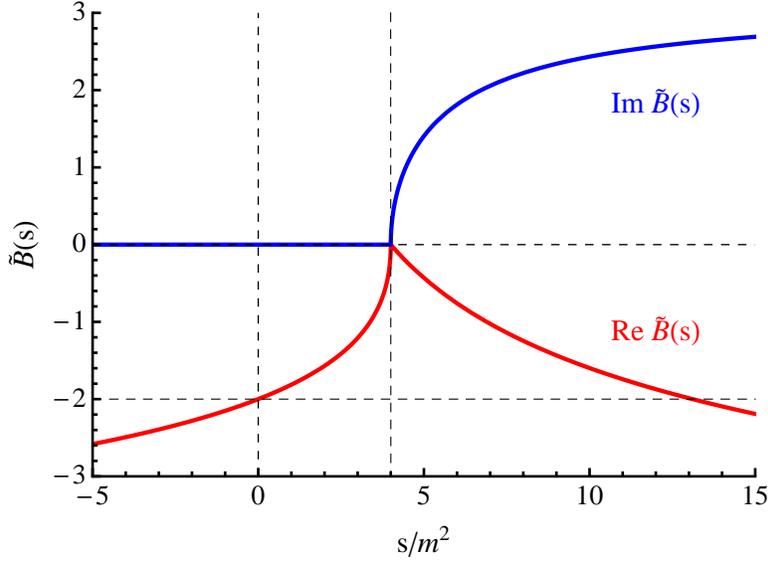}
\caption{Real and imaginary part of the renormalized one-loop correction to the four-point function of Eq.~(\ref{renB}), 
as function of $s$.}
\label{B}
\hspace{0.5cm}
\end{figure}

The resulting amplitudes are independent of the renormalization scale
$\mu$ which is achieved as usual by adding a counter term corresponding to 
the coupling constant renomalization. We define a renormalized coupling constant $\tilde \l$ by the equation
\beq
\tilde \lambda^{-1}(\mu^2)Ê= \tilde \l_0^{-1} + ÊL_\ve(\m^2) -2.
\label{renorm}
\eq
This gap equation allows to renormalize the calculated contribution to each order perturbatively.  
Using such renormalization, leads to the renormalized (subtracted) one-loop four-point function~:
\beq
\tilde B(s) \equiv (4 \pi)^2 \left[ B(s) - B(4 m^2) \right] = - \b(s)\,\ln\,\frac{1+\b(s)}{1-\b(s)} + i \p \b(s) \theta(s-4m^2),
\label{renB}
\eq
where we have also absorbed a factor $(4 \pi)^2$ in its definition.
We notice that our renormalization procedure is conveniently chosen so as to make a subtraction at threshold: $\tilde B(4 m^2) = 0$. 
We show the real and imaginary parts of the renormalized function $\tilde B(s)$ versus $s$ in Fig.~\ref{B}. 

\begin{figure}[ht]
\centering
\includegraphics[scale=0.8]{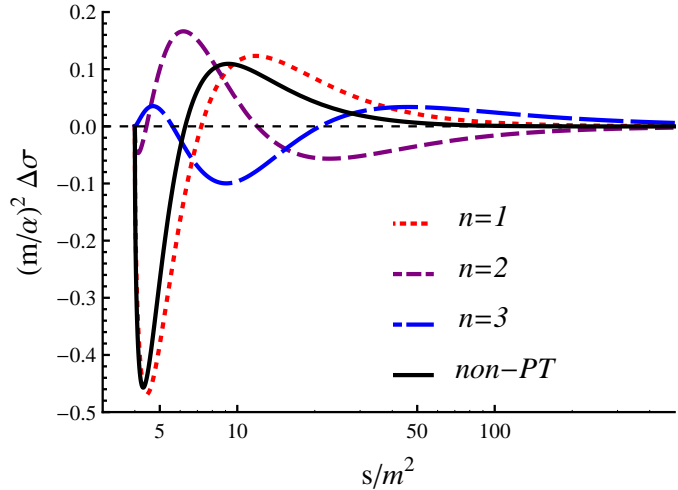}
\caption{The integrand of the sum rule Eq.~(\ref{Dsr}) for the scalar QED with four-particle self-interaction. The (red) fine-dashed line corresponds with the one-loop approximation, the (purple and blue) medium- and large-dashed lines correspond with  two- and three-loop approximations respectively, the (black) solid line corresponds with the non-perturbative sum of loop diagrams. The integrands are shown for $\tilde \l^{-1}= 4$.}
\label{nonPTsigma}
\end{figure}

The interference of two chain diagrams with total number of $(n-1)$ bubble loops  gives rise to a cross-section correction of the order 
$\mathcal{O}(\tilde{\l}^n)$. For the helicity-difference cross-section, which in the given case is equal to the helicity-$0$ cross section, we obtain 
as correction beyond the tree-level~: 
\beq
\begin{split}
\D\s^{(n)}(s)&=-\a^2  {\tilde \l}^{n} \frac{4 \p}{s} \b(s) \left\{\xi(s)\mathrm{Re}\left[ F(s) \tilde B^{n-1}(s)\right]
+ | F(s)|^2\sum\limits_{i=0}^{n-2} \tilde B^i(s)\[ \tilde B^{n-2-i}(s)\]^\ast\right\},
\label{PTcs}
\end{split}
\eq
where we used the notation
\beq
\xi(s)=  \frac{2 \, \left( 1 - \beta^2(s) \right)}{\b(s)}   \mathrm{arctanh}\, \beta(s) .
\eq
One can check explicitly that the expression of Eq.~(\ref{PTcs}) satisfies the helicity-difference sum rule exactly in each order of perturbation theory, i.e.
\beq
\mathrm{I}^{(n)}(\tilde \l) \equiv \int\limits_{4m^2}^\infty\dd s\,\frac{\D\s^{(n)}(s)}{s}=0. 
\label{Dsr}
\eq
In Fig.~\ref{nonPTsigma}, we plot the integrands of the sum rule at different orders of perturbation theory. One sees that in all cases the low- and high-energy contributions cancel.

It is well-known in quantum field theory that when trying to sum the different orders of perturbation theory, the resulting series is  mostly divergent and has at best a meaning as an asymptotic series, 
a notable exception being the case of theories which have the property of asymptotic freedom \cite{ Dyson:1952tj,Brezin:1976vw,Lipatov:1976ny}. 
As a result, the perturbation theory often can not determine the solution uniquely. If we are only interested in perturbative phenomena, it is not a problem to deal with it in the region of small coupling constants. However if we are interested in non-perturbative phenomena then we are 
faced with the problem to give a meaning to a divergent series. 

For example, the photon propagator calculated in the leading logarithmic approximation possesses an unphysical pole associated with negative invariant mass and negative probability, usually called the "Landau ghost" \cite{Landau:1954}. In QED, however, due to the smallness of the coupling constant the ghost appears at an extremely high energy scale, and the results at the energy scales accessible in experiment 
are not influenced by it. However, in contrast to the QED case, in the theory under consideration the coupling constant is not constrained to small values, and we are faced with the need to regularize our solution. The appearance of the Landau singularity is usually attributed to non-Borel-summability of the considered series, where some individual Feynman amplitudes are positive and grow like the factorial of the number of vertices producing a singularity on the real positive axis of the Borel transform of the perturbative series \cite{Gross:1974jv}. A similar situation arises also in the context of our model. We can easily see how non-Borel-summability manifests itself in the context of sum rules, by analyzing the contribution  from the high-energy region to the sum rule integral. In the example under study, the cross section Eq.~(\ref{PTcs}) in the $n$-bubble approximation 
behaves at large $s$ as~:
\beq
\D\s^{(n)}(s)\sim (-\tilde{\l})^n\frac{(\ln s/m^2)^{n-2}}{s}\, .
\eq
The contribution of the high-energy region to the sum rule integral can then be approximated by~:
\beq
\int\limits_{\L^2}^\infty\frac{\D\s^{(n)}(s)}{s}\dd s\sim\int\limits_{\L^2}^\infty (- \tilde \l)^n\frac{(\ln s/m^2)^{n-2}}{s^2}\dd s
=(- \tilde \l)^n(n-2)!\frac1{\L^2}\(1+\ln\frac{\L^2}{m^2}+\mathcal{O}(1/n)\) \,,
\eq
where $\L^2\gg m^2$.
As we see for negative $\tilde \lambda$ the high-energy contribution to the sum rule integral is positive definite and grows factorially with the order of perturbation theory, which amounts to the non-Borel summability of the series 
\beq
\mathrm{I}(\tilde \l) \equiv \sum\limits_{n=0}^\infty\mathrm{I}^{(n)}(\tilde \l)=\sum\limits_{n=0}^\infty\int\limits_{s_0}^\infty\frac{\D\s^{(n)}(s)}s\,\dd s \, .
\label{ser}
\eq
As a result we can consider $\sum_{n=0}^\infty\D\s^{(n)}(s)$ only in the sense of an asymptotic series. According to Poincar$\mathrm{\acute{e}}$ \cite{dingle}, a divergent series is an asymptotic expansion of a function $f(\l)$ if

\beq
\lim\limits_{|\l|\ra0}\[\frac1{\l^n}\left|f(\l)-\sum\limits_{k=0}^nf_k\l^k\right|\]=0,\qquad\mathrm{for}\; n\geqslant0.
\label{poincare}
\eq
This definition implies that an asymptotic series does not define a function uniquely. The expansion coefficients of a function of the type $e^{-1/\l}\sum_{k=0}^nf_k\l^k$ being identically zero, such a function can be added to $f(\l)$ without changing Eq.~(\ref{poincare})\cite{Kleinert:2001ax}. 

\begin{figure}[h]
\centering
  \includegraphics[width=9cm]{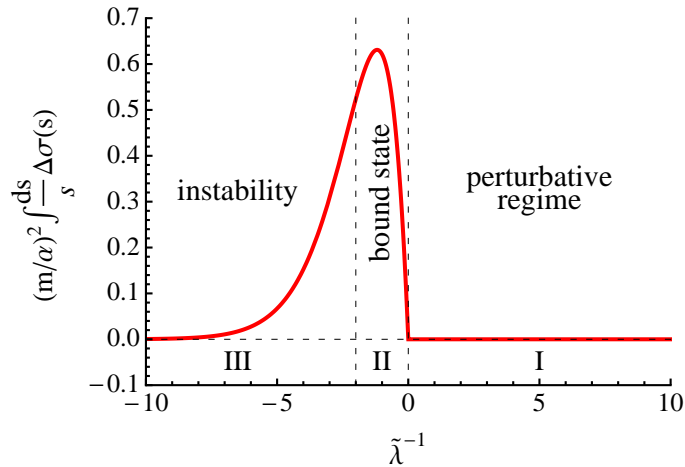}\\
  \caption{The dependence of the sum rule integral for the helicity difference cross section of the 
  $\gamma \gamma \to X$ process on the inverse coupling $\tilde \l^{-1}$.}
    \label{DSR}
\end{figure}

Thus it is natural to expect that in order to obtain the correct behavior of $I(\tilde \l)$ for negative $\tilde \l$, one has to modify a formal resummation  of the geometric series of renormalized bubble-chain corrections to the cross section in Eq.~(\ref{PTcs}), which is given by~: 
\beq
\D\s(s)=\sum\limits_{n=0}^{\infty}\D\s^{(n)}(s)= \D \s^{(\mathrm{tree})} - \a^2 \frac{4 \p}{s} \b(s) \left\{\xi(s)\mathrm{Re}\left[\frac{F(s)}{\tilde \l^{-1}- \tilde B(s)}\right] + 
\left|\frac{F(s)}{\tilde \l^{-1}- \tilde B(s)}\right|^2\right\},
\label{nPT}
\eq
where the tree-level cross section satisfies the sum rule by itself. 
As was discussed above, due to the non-Borel-summability of the series of Eq.~(\ref{ser}),  we do not have any reasons to expect that the sum rule integral 
will vanish for the resummed theory.  
In Fig.~\ref{DSR} we show the dependence of the sum rule integral for the cross section of Eq.~(\ref{nPT}) on the value of $\tilde \l$. 
We indeed notice from Fig.~\ref{DSR}  that the sum rule is only valid for positive values of $\tilde \l$ (denoted by region I), 
but is violated for negative values of $\tilde \l$ (regions II and III on Fig.~\ref{DSR}), showing that the naive procedure of the resummation 
is not applicable.  
In order to preserve validity of the sum rules beyond the region I, we need to find a way to evaluate the cross section correctly.
We will discuss the physical situation for the three regions of $\tilde \l$ in the following. 

\section{Discussion of the results}
\label{sec4}

\subsection*{Region I : convergent perturbative expansion}

Though the sum of Eq.~(\ref{ser}) is formally undetermined, we can still use a naive resummation at least in the region of positive $\tilde \l$, 
as one can see from Fig.~\ref{DSR}. For $\tilde \l^{-1}>0$ (region I) the series is alternating-sign, since $\mathrm{Re} \tilde B(s)<0$,
and one can expect the series to be resummable. 
Alternatively we can interpret such a resummation at the level of photon-photon-fusion amplitudes by summing up contributions of bubble-chain diagrams at different orders, which yields the amplitude~:
\beq
M_{++}(s, \theta)= M_{++}^{(tree)}(s, \theta)+  4 \pi \a \, \frac{2 F(s)}{\tilde \l^{-1}- \tilde B(s)}\, ,
\label{nPTa}
\eq
with tree-level amplitude given as in Eq.~(\ref{helampl}).
Squaring the amplitude of Eq.~(\ref{nPTa}) then yields the cross section of Eq.~(\ref{nPT}).
In the region I, the amplitude (\ref{nPTa}) has no poles for all complex values of $s$, and the series is conventionally convergent for all values of $s$. Thus the formal resummation of Eq.~(\ref{nPT}) satisfies the sum rule.

\subsection*{Region II~: Bound state and K-matrix pole}

Now we proceed to the discussion of the second region, denoted by region II on Fig.~\ref{DSR}. In the domain of $-2 < \tilde \l^{-1} <  0$ the sum rule is not valid for the naive cross section, given by Eq.~(\ref{nPT}). It is easy to see that the amplitude of Eq.~(\ref{nPTa}) acquires additional singularities below the  two-particle production threshold. Indeed from Fig.~\ref{B} we can see that the imaginary part of $\tilde B(s)$ vanishes for $s<4m^2$ and monotonically increases for $s>4m^2$. The real part satisfies the inequality
\beq
-\infty<\mathrm{Re} \tilde B(s) \leqslant 0,
\eq
for $0\leqslant s<\infty$, with the upper limit attained for $s=4m^2$. Thus we see that for $\tilde \l^{-1}<0$ there is always a $S$-matrix pole for $s<4m^2$. Above the two-particle threshold, there is a value of $s$ where $\tilde \l^{-1} =  {\mathrm {Re}} \tilde B(s)$, corresponding with a 
$K$-matrix pole as will be discussed below.  

We first discuss the pole below the threshold, which obviously corresponds to a bound state of the scalar pair. 
One notices from Fig.~\ref{B} that the mass of the bound state varies continuously from $M_B^2=4m^2$ for 
$\tilde \l^{-1} = 0$ to $M_B^2=0$ for $\tilde \l^{-1} = -2$ 
as we sweep over the bound state region on Fig. \ref{DSR}. In general, the position of the bound state pole is defined by the equation~:
\beq
\tilde \l^{-1}= \tilde B(M_B^2).
\eq 
It is important to note that the bound state singularity is not described by perturbation theory being essentially of non-perturbative nature.

\begin{figure}[h]
\centering
\includegraphics[scale=1]{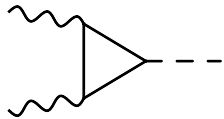}
\caption{The amplitude of the bound state production in the $\gamma \gamma$-fusion process.}
\label{ggb}
\end{figure}

The agreement with the sum rule can now be remedied by treating
the bound state as a new asymptotic state of the theory and thus including
the channel of its production in $\g\g$ collision, see Fig.~\ref{ggb}.
The corresponding contribution to the helicity-0 cross section is:
\beq
\s_0^{(\g\g\to B)} (s) = 4 \pi \a^2 \frac{g^2_{\mathrm{eff}}}{M_B^2}
| F(M_B^2)|^2\, \d( s-M_B^2)  ,
\eq
where the effective coupling of the bound state to $\phi \phi^\ast$ is 
found as the residue of the pole:
\beq
g^2_{\mathrm{eff}} =\frac{1}{| \tilde B^{\,\prime}(M_B^2) |}, \qquad 
 \tilde B^{\,\prime}(M_B^2)=\frac{d}{ds} \tilde B(s) \Big|_{s=M_B^2}.
\eq
It is not difficult to see now that its contribution to the sum rule integral:
\beq
\int_{0}^\infty \frac{ds}{s} \, \s_0^{(\g\g\to B)} (s) = 
4 \pi \a^2 \, \frac{g^2_{\mathrm{eff}}}{M_B^4}\,
| F(M_B^2)|^2 ,
\eq
exactly counter-balances the contribution of the $\g\g \to \phi\phi^\ast$ 
channel shown in Fig.~\ref{DSR}.
Thus the causality is restored in this region of $\tilde \l$.

\begin{figure}[h]
\centering
\includegraphics[scale=0.7]{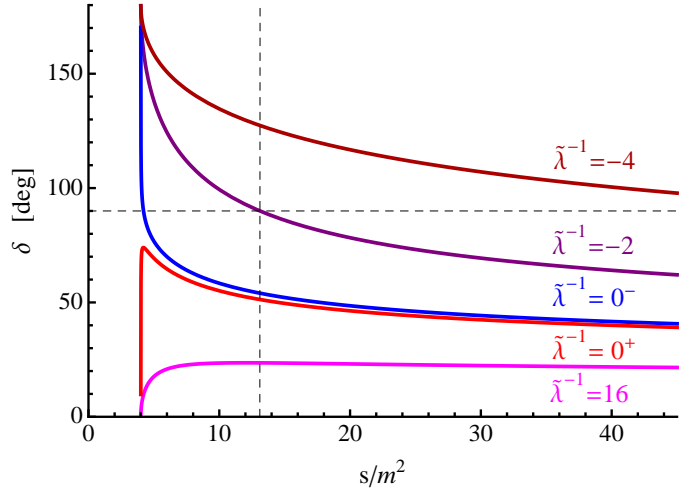}
\caption{Phase shift for different values of $\tilde \l$.}
\label{phsh}
\end{figure}

We now turn to the position of  the singularity above the two-particle threshold. 
To describe the elastic $\phi\phi$ scattering in $\phi^4$ theory, we only need to 
consider $S$-wave scattering in the bubble-chain approximation. 
The (dimensionless) elastic forward scattering amplitude $f(s)$ is expressed through a real phase shift $\d(s)$ 
as:
\beq
f(s)= e^{i\d(s)} \sin\d(s) ,
\eq
or through the $K$-matrix amplitude $K(s) \equiv \tan\d(s)$ as:
\beq
f(s) = \frac{1}{K^{-1}(s)- i } .
\eq
Since the imaginary part of the loop function is
\beq
\mathrm{Im} \,  B(s) = \p \b(s) \, \theta\(s-4m^2\),
\eq
for $s\geq 4m^2$ we can define the elastic amplitude as~:
\beq
f(s)= \pi \b(s) \frac{1}{\tilde \lambda^{-1} - \tilde B(s)} Ê=
\left(\frac{1}{\tilde \l \pi \b(s) } + \frac{2}{\p} \mathrm{arctanh} \b(s) ÊÊ- i \right)^{-1},
\eq
and hence
\beq
K^{-1} (s) = \frac{1}{\tilde \l \pi \b(s) } + \frac{2}{\pi} \mathrm{arctanh} \b(s). 
\eq
In Fig.~\ref{phsh} 
we show plots of the phase shift for different values of the coupling constant. 
Note that for negative $\tilde \l$ the phase-shift starts from $\pi $ which 
indicates the presence of one bound state. Also for negative $\tilde \l$
the phase shift crosses $\pi/2$ at $s>4m^2$ satisfying the following equation:
\beq
\tilde \l^{-1} = \mathrm{Re}\,  \tilde B(M_K^2).
\eq
This is the location of the $K$-matrix pole, corresponding with a scattering amplitude which 
becomes purely imaginary. Usually this behaviour is
attributed to a resonance. 
Since above the threshold, the imaginary part of $B(s)$ is not zero for all complex $s$ and 
$\tilde \l$ is real, the amplitude (\ref{nPTa}) does not possess a $S$-matrix pole for $s>4m^2$.
Hence there is no resonance associated with this $K$-matrix pole.
Note that on the right side of region II on Fig.~\ref{DSR}, corresponding with $\tilde \lambda^{-1} = 0$,
the $K$-matrix pole merges with the bound state and is defined by the position $M^2_K=4m^2$.
When reaching the left side of region II on Fig.~\ref{DSR}, corresponding with $\tilde \lambda^{-1} = -2$, 
the mass is obtained from 
$\mathrm{Re} B(M^2_K) =0$, which implies $M^2_K \approx13.1m^2$.

\subsection*{Region III~: Ground state instability and tachyonic solution}

If $\tilde \l^{-1}$ becomes smaller than $-2$ (corresponding with region III on Fig.~\ref{DSR}), the binding energy of the bound state exceeds $2m$,  and the pole crosses the point $s=0$, moving into the unphysical region $s<0$, and producing a tachyonic solution 
(a pole with negative invariant mass). 
The occurrence of a pole for negative values of $s$ signals that the ground state of the theory is unstable~\cite{Coleman:1974jh,Abbott:1975bn}. 
We could formally still include this pole in the $\g\g$-production channel as an asymptotic state and add its contribution to the total cross section in the same way as we did for the bound state, which restores the validity of the sum rule. For small negative values of $\tilde \l$, i.e. $\tilde \l^{-1} \to - \infty$, corresponding with the left asymptotic edge of the region III in Fig. \ref{DSR}, the contribution of the tachyon pole is vanishingly small and the sum rule is satisfied approximately. This is consistent with the observations in other models, for example, in the leading logarithmic approximation in QED, where the position of the Landau pole appears at very large scales due to the smallness of the fine structure constant. The position of the tachyon pole in this limit is defined by 
\beq
M^2\sim -e^{-1/\tilde \l}, 
\eq
which shows explicitly the non-perturbative structure of this contribution. The contribution of this pole in this sum rule is asymptotically defined as
\beq
\D\s(s)\sim 1 / M^2 \sim -e^{1/\tilde \l}
\label{tachSR}
\eq
and vanishes when $\tilde \l^{-1} \ra - \infty$.
However such a procedure has a number of inconsistencies. Being required by the sum rule such extra contribution spoils the general principles of field theory. The cross section of Eq.~(\ref{tachSR}) is negative, thus indicating that the appearance of a tachyon ghost state contradicts the usual quantum mechanical probabilistic interpretation and spoils unitarity. Moreover, the delta function is located at a space-like squared four momentum $-M^2$ indicating the occurrence of a tachyonic instability and spoiling the analyticity principle. 
All these facts show that the bubble-chain approximation is essentially inconsistent in region III of the coupling constant. 
For the full theory, such unphysical state will not appear among the exact eigenstates.

\section{Conclusions}
\label{sec5}

We have studied consequences of causality constraints imposed by 
a recently established sum rule for the forward light-by-light scattering process 
within the $\phi^4$ scalar quantum field theory. 
Within this theory, we verified the sum rule at the one-loop level as well as to all orders within the bubble-chain approximation. 
Furthermore, we have performed a resummation of bubble graphs. We have 
demonstrated that depending on the value of the renormalized self-interaction coupling constant 
of the $\phi^4$ theory, three different regimes emerge. 
In a first regime, the perturbative series is convergent and the sum rule as calculated from the continuum states in the theory holds exactly. 
In a second regime for the renormalized coupling, the resummed amplitude acquires additional singularities: a dynamically generated 
bound state below the two-particle production threshold and a $K$-matrix pole above the two-particle production. 
It was shown that when evaluating the light-by-light sum rule, the bound state contribution exactly cancels the continuum contribution, so 
as to verify the sum rule. Furthermore, we found a third regime of the renormalized coupling where a tachyonic solution with negative invariant mass appears, signaling that in this regime the vacuum is unstable and that the considered 
bubble-chain resummation is essentially inconsistent. 

The results within the considered model relativistic quantum field theory have demonstrated that light-by-light scattering sum rules provide a very powerful tool 
to quantitatively connect dynamically generated bound states with the continuum region of the theory. As such this can be a first step, to apply such a tool e.g. to the study of mesons produced in the $\gamma \gamma$-fusion process presently under study at different $e^+e^-$ collider facilities. 

\section*{Acknowledgements}

This work was supported by the Deutsche Forschungsgemeinschaft DFG in part through the Collaborative 
Research Center ``The Low-Energy Frontier of the Standard Model" (SFB 1044), 
in part through the graduate school Graduate School ``Symmetry Breaking in Fundamental Interactions" 
(DFG/GRK 1581), and in part through 
the Cluster of Excellence "Precision Physics, Fundamental Interactions and Structure of Matter" (PRISMA).

\end{document}